\documentclass[11pt]{article}

\usepackage{fullpage}
\usepackage{tikz}
\usetikzlibrary{arrows,automata,shapes,decorations,calc,matrix,decorations.pathmorphing}
\usepackage{graphicx}
\usepackage{caption}
\usepackage{subcaption}
\usepackage{xcolor}

\usepackage{amssymb,amsfonts,amsmath,amsthm}

\usepackage{authblk}
\usepackage{blindtext}
\usepackage{lineno}

\newtheorem{theorem}{Theorem}
\theoremstyle{definition}
\newtheorem{lemma}{Lemma}
\newtheorem*{claim}{Claim}
\newtheorem{corollary}{Corollary}

\def\isf{\operatorname{isf}}
\def\d{d}
\def\din{d_{in}}

\def\Ev{\mathcal{E}}

\def\eps{\varepsilon}

\def\epdelbar{{\widehat{\operatorname{ep}}^\delta}}
\def\epd{\operatorname{ep}^{\operatorname{dB}}}
\def\ep{\operatorname{ep}^{\operatorname{Bd}}}

\def\fpsign{\rho}

\def\fp{\fpsign^{Bd}}
\def\fpd{\fpsign^{dB}}
\def\fpdel{\fpsign^{\delta}}
\def\epdel{1-\fpsign^{\delta}}
\def\fpdelbar{\widehat{\fpsign}^{\delta}}

\def\sne#1{\medskip\noindent\textbf{#1.}\ }

\newcommand{\old}[1]{{\color{pink} }}

\newcommand\numberthis{\addtocounter{equation}{1}\tag{\theequation}}

\def\lemmadb{
\begin{lemma}\label{lemma} 
Fix $r>1$ and let $G$ be a graph (possibly with directed and/or weighted edges) with average out-degree $\d$. Then
$$\fpd(G,r)\le  \frac{\d\cdot r}{\d\cdot r+\d+r-1}.
$$
\end{lemma}
}

\def\thmTransient{
\begin{theorem}[All dB amplifiers are transient]\label{thm:transient}
Fix a non-complete graph $G_N$ (possibly with directed and/or weighted edges). Then there exists $r^\star>1$ such that for all $r>r^\star$ we have $\fpd(G_N,r)< \fpd(K_N,r)$, where $K_N$ is the complete graph on $N$ vertices. In particular, we can take $r^\star=2N^2$.
\end{theorem}
}

\def\thmBounded{
\begin{theorem}[All dB amplifiers are bounded]\label{thm:bounded}
Fix $r>1$. Then for any graph $G_N$ (possibly with directed and/or weighted edges) we have $\fpd(G_N,r)\le 1-\frac1{r+1}$. 
\end{theorem}
}

\def\lemmadelta{
\begin{lemma}\label{lemma2} Fix $r>1$ and let $G$ be a graph (possibly with directed and/or weighted edges) with average out-degree $\d$. Then
$$\epdel(G,r)\ge  \frac{1}{1+  \frac{\d r}{d+r-1} +\frac{1-\delta}\delta\cdot \frac{Nr}{N+r-1}  }.
$$
\end{lemma}
}

\def\thmMixedTransient{
\begin{theorem}[All $\delta$-dB amplifiers are transient]\label{thm:mixedtransient}
Fix a non-complete graph $G$ on $N$ vertices (possibly with directed and/or weighted edges) and $\delta\in(0,1]$. Then there exists $r^\star>1$ such that for all $r>r^\star$ we have $\fpdel(G,r) < \fpdelbar(K_N,r)$, where $K_N$ is the complete graph on $N$ vertices.
\end{theorem}
}

\def\thmMixed{
\begin{theorem}[All $\delta$-dB amplifiers are at most linear]\label{thm:mixed}
Fix $r>1$ and $\delta\in(0,1]$. Then for any graph $G$ (possibly with directed and/or weighted edges) we have $\fpdel(G,r)\le 1- \frac 1{(r/\delta) +1}$.
\end{theorem}
}

\def\Title{Limits on amplifiers of natural selection under death-Birth updating}

\def\Title{Limits on amplifiers of natural selection under death-Birth updating}

\begin{document}

\title{{
\Title
}}
\date{}

\author[a, 1]{Josef\! Tkadlec}
\author[b, 1]{Andreas\! Pavlogiannis}
\author[a]{Krishnendu\! Chatterjee}
\author[c, $\star$]{Martin\! A.\! Nowak}
\affil[a]{IST Austria, A-3400 Klosterneuburg, Austria}
\affil[b]{Lab for Automated Reasoning and Analysis, EPFL, CH-1015 Lausanne, Switzerland}
\affil[c]{Program for Evolutionary Dynamics, Department of Organismic and Evolutionary Biology, Department of Mathematics, Harvard University, Cambridge, MA 02138, USA}
\affil[1]{J.T. and A.P. contributed equally to this work.}
\affil[$\star$]{To whom correspondence should be addressed. E-mail: martin\_nowak@harvard.edu}


\maketitle

\tableofcontents

\section*{Abstract}
The fixation probability of a single mutant invading a population of residents is among the most widely-studied quantities in evolutionary dynamics.
Amplifiers of natural selection are population structures that increase the fixation probability of advantageous mutants, compared to well-mixed populations.
Extensive studies have shown that many amplifiers exist for the Birth-death Moran process,
some of them substantially increasing the fixation probability or even guaranteeing fixation in the limit of large population size.
On the other hand, no amplifiers are known for the death-Birth Moran process, and computer-assisted exhaustive searches have failed to discover amplification.
In this work we resolve this disparity, by showing that
any amplification under death-Birth updating is necessarily \emph{bounded} and \emph{transient}.
Our boundedness result states that even if a population structure does amplify selection, the resulting fixation probability is close to that of the well-mixed population.
Our transience result states that for any population structure there exists a threshold $r^*$ such that the population structure ceases to amplify selection if the mutant fitness advantage $r$ is larger than $r^\star$.
 Finally, we also extend the above results to $\delta$-death-Birth updating, which is a combination of Birth-death and death-Birth updating.
 On the positive side, we identify population structures that maintain amplification for a wide range of values $r$ and $\delta$.
 These results demonstrate that amplification of natural selection depends on the specific mechanisms of the evolutionary process.


\section{Introduction}
The evolutionary rate of populations is determined by their ability to accumulate advantageous mutations~\cite{Kimura68,Ewens04,Nowak2006,Desai07,McCandlish15}.
Once a new mutant has been randomly generated in a population, its fate is governed by the dynamics of natural selection and random drift.
The most important quantity in this process is the \emph{fixation probability} which is the probability that the invading mutant fixates in the population as opposed to being swept away.
A classical mathematical framework for rigorous study of the mutant spread is the discrete-time Moran process~\cite{Moran1962}.
Given a population of $N$ individuals, at each time step, 
(1)~an individual is chosen randomly for reproduction proportionally to its fitness and
(2)~an individual dies uniformly at random;
then the offspring of the reproducing individual replaces the dead individual, and the population size remains constant.

\begin{figure}[ht] 
  \centering
   \includegraphics[width=1\linewidth]{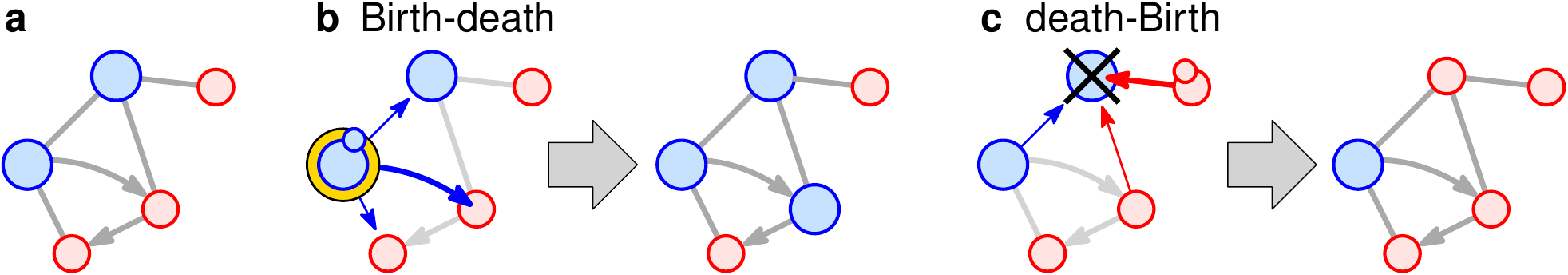}
\caption{ \textbf{Moran process on graphs.}
\textbf{a,} The spatial structure is represented by a graph. Each vertex represents a site and is occupied either by a resident (red) with fitness $1$ or by a mutant (blue) with relative fitness $r>1$. Each edge can be one-way (arrow) or two-way. 
\textbf{b,} In each step of the Birth-death process, one individual is sampled for reproduction proportionally to fitness, and then its offspring replaces a random neighbor. 
\textbf{c,} In each step of the death-Birth process, a random individual dies and then it is replaced by a neighbor sampled proportionally to fitness.
}
\label{fig:fig1}
\end{figure}

Many evolutionary properties are affected by the spatial arrangement of the population~~\cite{Slatkin81,Nowak1992,durrett1994stochastic,Whitlock2003,Hauert04,komarova2006spatial,Houchmandzadeh2011,Frean2013,komarova2014complex}.
Evolutionary graph theory represents population structure of size $N$ by a graph (network) $G_N$~\cite{lieberman2005,Broom2008,Broom2011,shakarian12,Debarre14,Allen2017}: 
each individual occupies a vertex, and neighboring vertices mark sites of spatial proximity (see Fig.~\ref{fig:fig1}a).
Mutant spread must respect the structure, in that the offspring of a reproducing individual in one vertex can only move to a neighboring vertex.
The Moran process on graphs has two distinct variants:
\begin{itemize}
\item In the \emph{Birth-death} Moran process, the death event is conditioned on the Birth event. That is, first an individual is chosen for reproduction and then its offspring replaces a random neighbor (see Fig.~\ref{fig:fig1}b).
\item In the \emph{death-Birth} Moran process, the Birth event is conditioned on the death event. That is, first an individual is chosen for death and then its neighbors compete to fill the vacancy with their offspring (see Fig.~\ref{fig:fig1}c).
\end{itemize}

The fixation probability of the invading mutant is a function of its fitness $r$, as well as the graph $G_N$.
In alignment with most of the literature, we focus on advantageous mutants, where $r>1$.

The well-mixed population of size $N$ is represented by a complete graph $K_N$.
In the Birth-death Moran process, the fixation probability in the well-mixed population is 
$\fp(K_N, r)=(1-1/r)/(1-1/r^{N})$~\cite{Nowak2006}.
Under death-Birth updating, the fixation probability is $\fpd(K_N, r)=(1-1/N)\cdot (1-1/r)/(1-1/r^{N-1})$~\cite{kaveh15duality}.
Specifically, as $N\to\infty$, both the expressions converge to $1-1/r$.

Amplifiers of natural selection are graphs that increase the fixation probability of the advantageous mutants compared to the well-mixed population~\cite{ACN15,lieberman2005}.
Under Birth-death updating, many amplifying families of graphs have been constructed, such as the Star graph~\cite{Chalub16,broom2011stars,askari2015analytical}, the Complete Bipartite graph~\cite{Monk2014} and the Comet graph~\cite{beatingstar},
as well as families that guarantee fixation in the limit of large population size~\cite{lieberman2005,Giakkoupis16,Galanis17,pavlogiannis2018construction,Goldberg19}.
Extensive computer simulations on small populations have also shown that many graphs have amplifying properties~\cite{Hindersin2014,HT15,Tkadlec19}.
While the above results hold for the Birth-death Moran process, no amplifiers are known for the death-Birth Moran process,
and computer-assisted search has found that, under death-Birth updating, most small graphs suppress the fixation probability rather than amplifying it~\cite{HT15}.

Here we prove two negative results on the existence of amplifiers under death-Birth updating.
Our first result states that the fixation probability in any graph is bounded by $1-1/(r+1)$.
Hence, even if amplifiers do exist, they can provide only limited amplification.
In particular, there are no families of graphs that would guarantee fixation in the limit of large population size.
Our second result states that for any graph $G_N$, there exists a threshold $r^*$ such that for all $r\geq r^*$,
the fixation probability is bounded by $\fpd(r, K_N)$.
Hence, even if some graphs amplify for certain values of $r$, their amplifying property is necessarily transient, and lost when the mutant fitness advantage $r$ becomes large enough.
We note that a companion work~\cite{BEN} identifies transient amplifiers among graphs that have weighted edges.
Finally, we also study the mixed $\delta$-death-Birth Moran process, for $\delta\in [0,1]$, under which
death-Birth and Birth-death updates happen with rate $\delta$ and $1-\delta$, respectively~\cite{Zukewich13}.
We establish analogous negative results for mixed $\delta$-updating, for any fixed $\delta>0$.
Note that as $\delta$ vanishes ($\delta\to 0$), we approach (pure) Birth-death Moran process for which both universal and super amplifiers exist.
We find that some of those amplifiers are less sensitive to variations in $\delta$ than other.
In particular, certain bipartite structures achieve transient amplification for $\delta$ as big as $0.5$.


\section{Model}

\subsection*{The Moran process on graphs}
In evolutionary graph theory, a population structure has traditionally been represented by a graph $G_N=(V,E)$,
where $V$ is the set of $N$ vertices representing sites and $E\subseteq V\times V$ is the set of edges representing neighborships between the sites.
We say that $G_N$ is \textit{undirected} when all edges are two-way, that is, $(v,u)$ is an edge whenever $(u,v)$ is.
Since the focus of this work is on death-Birth updating,
we require that there are no self-loops in $G_N$ (that is, $(u,u)$ is never an edge).
More generally, a population structure can be represented by a \textit{weighted} graph. In that case, every edge $(u,v)$ is assigned a weight $w_{u,v}\in[0,1]$
which indicates the strength of interaction from site $u$ to site $v$.
In full generality, we allow for non-symmetric weights (that is, possibly $w_{u,v}\neq w_{v,u}$).
The family of \textit{unweighted} graphs is recovered when we insist that all edges have weight 1.
Even though our primary focus is on unweighted graphs, our results apply to weighted graphs too.
A population of $N$ residents inhabits the graph $G_N$ with a single individual occupying each of the vertices of $G_N$.

In the beginning of the Moran process, one vertex is chosen uniformly at random to host the initial mutant.
The mutant has a fitness advantage $r>1$, whereas each of the residents has fitness normalized to $1$.
We denote by $f(u)$ the fitness of the individual occupying the vertex $u$.
From that point on, the process proceeds in discrete time steps, according to one of the two variants of updating:
\begin{enumerate}
\item Under \emph{death-Birth (dB)} updating, first an individual is selected to die uniformly at random.
This leaves a vacancy in the corresponding vertex $v$ of $G_N$, and the neighbors of $v$ compete to fill it.
Specifically, every neighbor $u$ of $v$ is chosen for reproduction with probability proportional to $f(u)\cdot w_{u,v}$,
and the selected individual places a copy of itself on $v$.
\item Under \emph{Birth-death (Bd)} updating, first an individual is selected to reproduce with probability proportional to its fitness, that is, with probability proportional to $f(u)$ for the individual occupying the vertex $u$.
The offspring of $u$ then replaces a random neighbor $v$ of $u$ with probability proportional to $w_{u,v}$.
\end{enumerate}
We also consider a combination of dB and Bd updating, which yields the mixed $\delta$-death-Birth Moran process.
\begin{enumerate}
\item[3] Under \emph{$\delta$-death-Birth ($\delta$-dB)} updating, for $\delta\in[0,1]$, in each step, the Moran process follows a dB update with probability $\delta$, and a Bd update with probability $1-\delta$.
In this notation, $\delta=1$ corresponds to pure dB updating, and $\delta=0$ corresponds to pure Bd updating.
\end{enumerate}
We only consider strongly connected graphs for which, with probability 1 in the long run, the Moran process leads either to the fixation of the mutant in the population (all vertices are eventually occupied by mutants) or to the extinction of the mutant (all vertices are eventually occupied by residents).
We denote by $\fpd(G_N, r)$, $\fp(G_N, r)$ and $\fpdel(G_N, r)$ the fixation probability under dB, Bd and $\delta$-dB updating, respectively.

\subsection*{Amplifiers}

The well-mixed population is modelled by the undirected complete graph $K_N$.
The fixation probability on $K_N$ under Bd updating is~\cite{Nowak2006}
\begin{align*}
\fp(K_N, r)=\frac{1-1/r}{1-1/r^{N}}. \numberthis\label{eq:well_mixed_bd}
\end{align*}
Similarly, the fixation probability on $K_N$ under dB updating is~\cite{kaveh15duality}
\begin{align*}
\fpd(K_N, r)=\left(1-\frac{1}{N}\right)\cdot  \frac{1-1/r}{1-1/r^{N-1}}. \numberthis \label{eq:well_mixed_db}
\end{align*}
Specifically, as $N\to\infty$, both the expressions converge to $1-1/r$.

Population structure can affect the fixation probability of advantageous mutants.
Given $r>1$, a graph $G_N$ is a Bd (resp., dB) \textit{$r$-amplifier} if 
$\fp(G_N, r) > \fp(K_N, r)$
(resp., $\fpd(G_N, r) > \fpd(K_N, r)$).
In words, a graph $G_N$ is an $r$-amplfiier the fixation probability of mutants with fitness advantage $r$ on $G_N$ is larger than the fixation probability of the same mutants on the well-mixed population.
If $G_N$ is an $r$-amplifier for all $r>1$, then we say that $G_N$ is a \textit{universal amplifier}.
(In the earlier literature, the word ``amplifier'' had typically been used to mean ``universal amplifier''.)
If $G_N$ is an $r$-amplifier for only a limited range of $r$-values, that is, there exists a threshold value $r^\star$ such that $G_N$ does not amplify for any $r>r^\star$, we say that $G_N$ is a \textit{transient amplifier}.
(Note that, in principle, an amplifier could be neither universal nor transient -- it could indefinitely alternate between amplifying and suppressing as $r$ grows larger.)

A notorious example of a universal Bd amplifier is a Star graph $S_N$ (of any fixed size $N\ge 3$) which consists of one central vertex connected to each of the $N-1$ surrounding leaf vertices. As $N\to\infty$, the fixation probability of a mutant with fitness advantage $r$ on $S_N$ converges to $\fp(S_N,r)\to_{N\to\infty} 1-1/r^2$. Effectively, the Star-like population structure rescales the fitness of the mutant from $r$ to $r^2$.
Under Bd updating, there also exist population structures that amplify for only some values of $r>1$~\cite{cuesta2018evolutionary}. 

\begin{figure}[h] 
  \centering
   \includegraphics[width=1\linewidth]{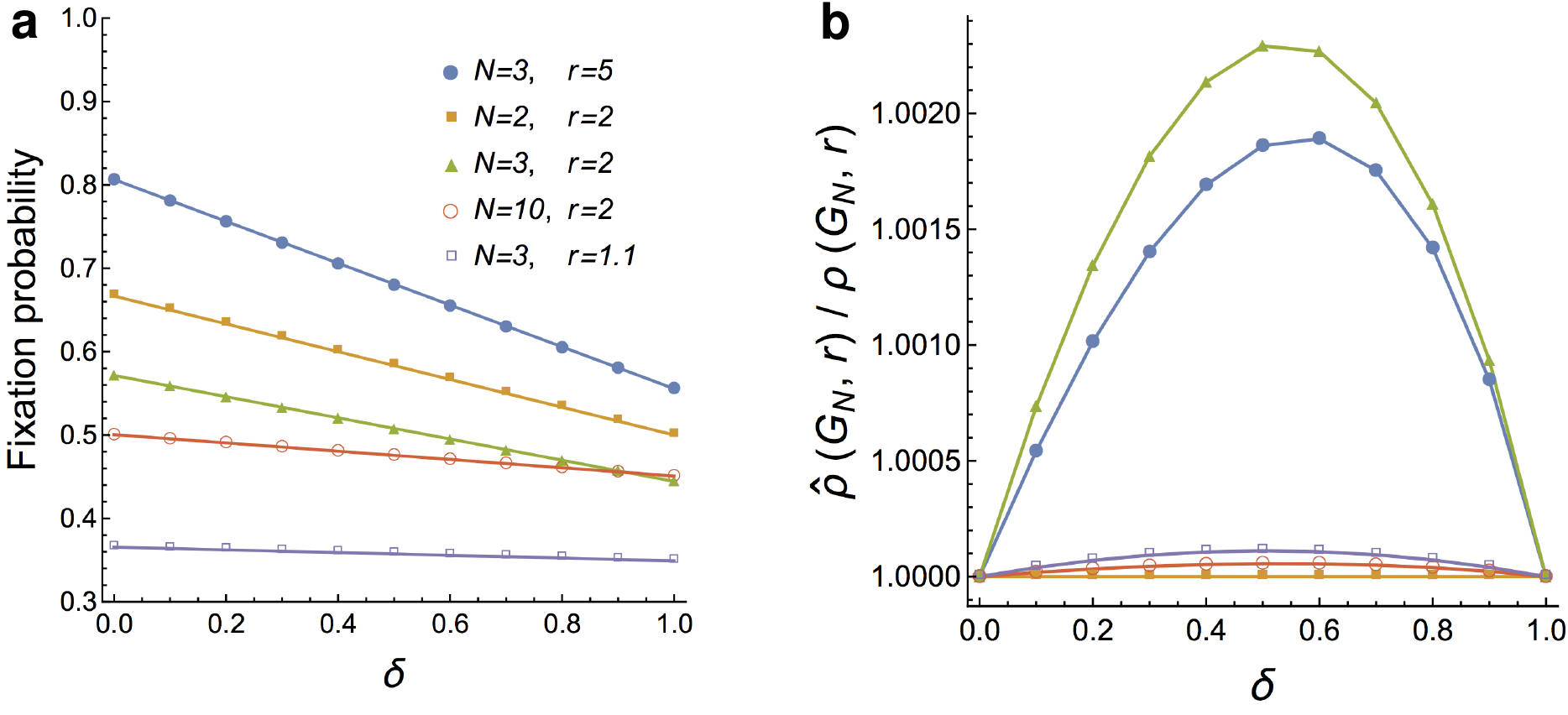}
\caption{ \textbf{Linear interpolation for $\delta$-dB updating.}
 On a complete graph $K_N$, the fixation probability $\fpdel(K_N,r)$ under $\delta$-dB updating is essentially indistinguishable from the linear interpolation $\fpdelbar(K_N,r)$ between fixation probability under pure dB and pure Bd updating.
\textbf{a,} The $x$-axis shows $\delta\in[0,1]$, the $y$-axis shows the fixation probability $\fpdel(K_N,r)$ (marks) and the linear interpolation $\fpdelbar(K_N,r)$ (lines) for several pairs $(N,r)$. The marks lie almost exactly on the lines.
\textbf{b,} The ratio $\fpdelbar(K_N,r) / \fpdel(K_N,r)$ is well within 1 \%, typically even within 0.1 \% of 1. The interpolation is exact for $N=2$.
}
\label{fig:fig2}
\end{figure}

Although the fixation probability on $K_N$ is known for dB and Bd updating, 
there is no known formula for $\fpdel(K_N,r)$.
We computed $\fpdel(K_N,r)$ numerically for various values of $N$ and $r$ and we observed that $\fpdel(K_N,r)$ is essentially indistinguishable from the linear interpolation 
\begin{align*}
\fpdelbar(K_N,r)= \delta\cdot \fpd(K_N,r) + (1-\delta)\cdot \fp(K_N,r)\numberthis\label{eq:well_mixed_delta}
\end{align*}
between $\fp(K_N,r)$ and $\fpd(K_N,r)$ (see Fig.~\ref{fig:fig2}a).
In fact, the ratio $\fpdelbar(K_N,r)/\fpdel(K_N,r)$ appears to be well within 1 \% of 1, and most of the time even within 0.01 \% of 1 (see Fig.~\ref{fig:fig2}b).
Therefore, in $\delta$-dB updating we use $\fpdelbar(K_N,r)$ as the baseline comparison,
and say that a graph $G_N$ is a $\delta$-dB $r$-amplifier if $\fpdel(G_N, r)>\fpdelbar(K_N,r)$.

\subsection*{Implied scale of fitness}
\begin{figure}[ht] 
  \centering
   \includegraphics[width=1\linewidth]{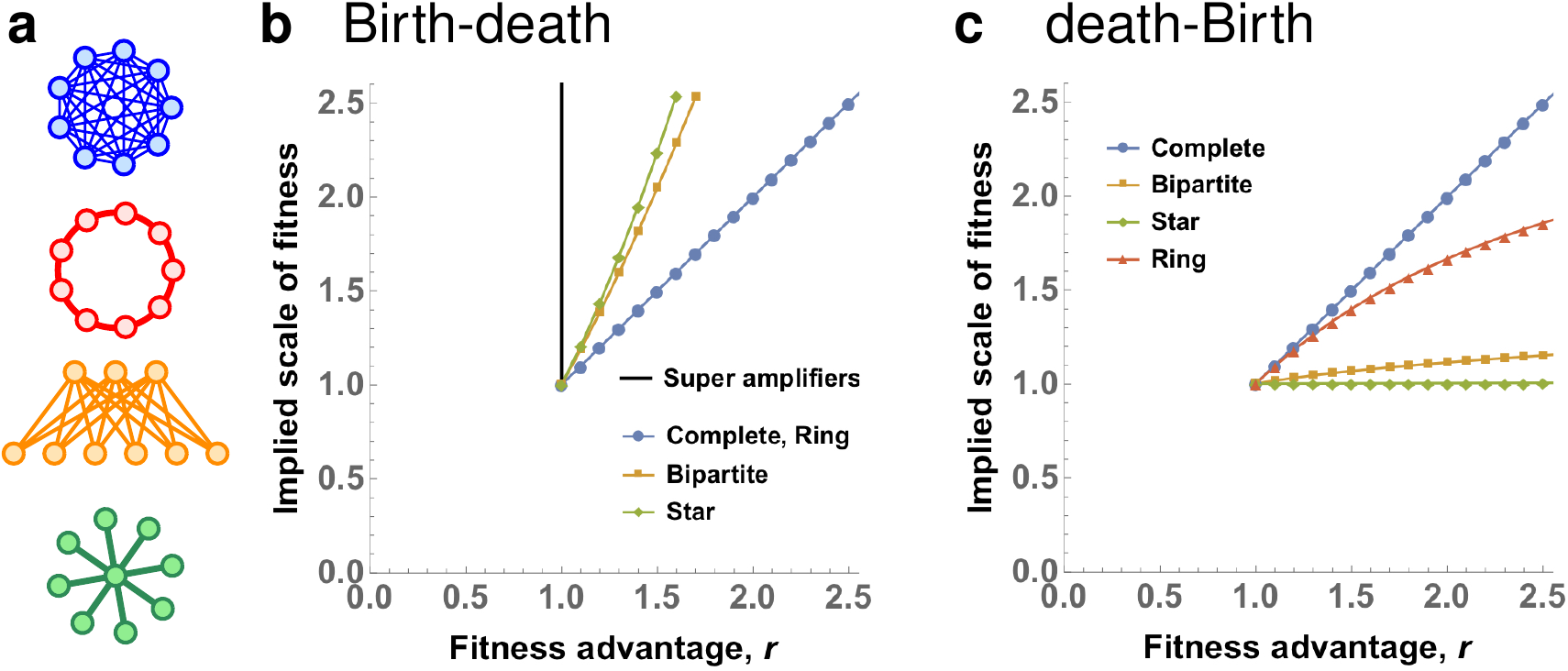}
\caption{ \textbf{Implied scale of fitness.}
The implied scale of fitness for several graph families.
\textbf{a,} Complete graphs $K_N$, Ring graphs $R_N$, complete Bipartite graphs $B_{\sqrt N,N-\sqrt N}$ and Star graphs $S_N$.
\textbf{b,} Under Birth-death updating, the Star graphs and the Bipartite graphs are quadratic amplifiers, whereas the Ring graphs are equivalent to Complete graphs. There also exist super amplifiers that guarantee fixation with probability 1 for any $r>1$. (To model the limit $N\to\infty$ we show values for $N=400$.)
\textbf{c,} Under death-Birth updating, none of Bipartite graphs, Star graphs or Ring graphs amplify selection.
}
\label{fig:figisf}
\end{figure}

We are typically interested in the fixation probability when the population size is large.
This leads us to the study of families of graphs $\{G_N\}_{N=1}^{\infty}$ of increasing population size,
the fixation probability of which is taken in the limit of $N\to\infty$.
Graph families can be classified by amplification strength.
Given such a family, the implied scale of fitness for that family~\cite{ACN15} is a function $\isf(r)$ such that
\begin{align*}
\lim\inf_{N\to \infty}\fpd(G_N,r) = 1-1/\isf(r)
\end{align*}
Specifically, for the family of complete graphs $K_N$ we have $\isf(r)=r$, under both dB updating and Bd updating.
We say that the family is (at most) a bounded amplifier if $\isf(r)\le r+c_0$ for some constant $c_0$.
We say that the family is (at least) a linear amplifier if $\isf(r)\ge c_1 r +c_0 $ for some constants $c_1>1$, $c_0$.
We say that the family is (at least) a quadratic amplifier if $\isf(r)\ge c_2 r^2 + c_1r+c_0$ for some constants $c_2>0$, $c_1$, $c_0$. For instance, Star graphs are quadratic amplifiers under Bd updating~\cite{Nowak2006}, however they do not amplify under dB updating~\cite{kaveh15duality}.
Finally, the family is a super amplifier if $\isf(r)=\infty$ for any $r>1$. That is, for any $r>1$ we have $\fpd(G_N,r)\to_{N\to\infty}=1$ and hence fixation is guaranteed in the limit of large population size (see Fig.~\ref{fig:figisf}).
The above definitions carry naturally to the $\delta$-dB Moran process, where the implied scale of fitness is defined such that
\begin{align*}
\lim\inf_{N\to \infty}\fpdel(G_N,r) = 1-1/\isf(r)
\end{align*}

\subsection*{Questions}

For the Bd Moran process, various results on amplifiers exist. The Star graph is a prominent example of a graph that is a quadratic amplifier for any $r>1$~\cite{Chalub16,broom2011stars,Monk2014,askari2015analytical,beatingstar} and there exist super amplifiers, that is, families of graphs that guarantee fixation in the limit of large population size, for any fixed $r>1$~\cite{lieberman2005,Giakkoupis16,Galanis17,pavlogiannis2018construction,Goldberg19}.
 Furthermore, computer simulations on small populations have shown that many small graphs have amplifying properties~\cite{Hindersin2014,HT15,Tkadlec19}.
Given the vast literature on results under Bd updating, the following questions arise naturally.
\begin{enumerate}
\item[Q1:] Do there exist universal amplifiers for the dB Moran process?
\item[Q2:] Do there exist families that are amplifying for the dB Moran process?
More specifically, do there exist linear, quadratic, or even super amplifiers?
\end{enumerate}
The first question is concerned with small populations, and asks for a graph that amplifies for all $r>1$.
The second question asks for amplification in the limit of large populations.

\section{Results}

Here we establish some useful observations about the dB Moran process, and then answer questions Q1 and Q2. 

First, consider the dB Moran process on any (fixed) graph $G_N$. 
The fixation probability can be bounded from above in terms of the number of neighbors of the vertex where the initial mutant has appeared.
As a simple example, consider that $G_N$ is unweighted and undirected, and each vertex has precisely $d$ neighbors (e.g. a square lattice where $d=4$). Denote by $v$ the vertex that hosts the initial mutant. 
We observe that if $v$ is selected for death before any of its $d$ neighbors then the mutants have just gone extinct. 
Since this event has probability $1/(d+1)$, the fixation probability is at most $1-1/(d+1) = d/(d+1)$, regardless of $r$. 
A more refined version of this argument, which also accounts for arbitrary graphs, yields the following stronger bound.

\begin{lemma}\label{lem:bound} 
Fix $r>1$ and let $G_N$ be a graph (possibly with directed and/or weighted edges) with average out-degree $\d$. Then
$$\fpd(G_N,r)\le  \frac{\d\cdot r}{\d\cdot r+\d+r-1}.
$$
\end{lemma}

For large enough $r$ and small (fixed) $\d$, the bound of Lemma~\ref{lem:bound} coincides with the bound we obtained with our sketchy argument above.
Observe that even when $r\to \infty$, the lemma yields an upper-bound on the fixation probability that is strictly less than 1.
On the other hand, under Bd updating, the fixation probability tends to $1$ as $r\to \infty$, regardless of the graph.
Hence we have the following corollary, which states that for all population structures, Bd updating favors fixation more than dB updating, provided that the fitness advantage is large enough.

\begin{corollary}\label{cor:dB_Bd}
For any graph $G_N$, there exists some $r^\star$, such that for all $r> r^\star$, we have
$\fp(G_N, r)>\fpd(G_N, r)$.
\end{corollary}

%

\subsection*{Amplifiers of the dB Moran process}
Here we answer the two questions Q1, Q2.
We start with Q1 which asks for the existence of universal amplifiers under dB updating.
We show the following theorem.

\begin{theorem}[All dB amplifiers are transient]\label{thm:transient}
Fix a non-complete graph $G_N$ (possibly with directed and/or weighted edges). Then there exists $r^\star>1$ such that for all $r>r^\star$ we have $\fpd(G_N,r)< \fpd(K_N,r)$, where $K_N$ is the complete graph on $N$ vertices. In particular, we can take $r^\star=2N^2$.
\end{theorem}
Since our baseline for amplifiers is the complete graph $K_N$,
Theorem~\ref{thm:transient} implies that, under dB updating, every (unweighted) graph is, at best, a transient amplifier.
Moreover, the only graph that may be a universal (that is, non-transient) amplifier is a weighted version of the complete graph $K_N$.
This is in sharp contrast to Bd updating, for which universal amplifiers exist (e.g., the Star graph~\cite{Monk2014}).

To sketch the intuition behind Theorem~\ref{thm:transient}, consider again, for simplicity, our toy example of an unweighted undirected graph $G_N$ where each vertex has precisely $d$ neighbors. Then the fixation probability is at most $d/(d+1)$, regardless of $r$. On the other hand, equation~\ref{eq:well_mixed_db} implies that the fixation probability on a complete graph tends to $1-1/N$ as $r\to \infty$. If $d<N-1$, then $1-1/N$ is strictly more than $d/(d+1)$, hence the graph $G_N$ ceases to amplify in the limit $r\to\infty$. In the proof, we use Lemma~\ref{lem:bound} which applies to possibly weighted, directed, and/or non-regular graphs and which yields an explicit bound on the threshold $r$-value $r^\star\le 2N^2$.

Second, we turn our attention to question Q2, which asks for the existence of strong amplifying families. We establish the following theorem, which answers Q2 in negative.
\begin{theorem}[All dB amplifiers are bounded]\label{thm:bounded}
Fix $r>1$. Then for any graph $G_N$ (possibly with directed and/or weighted edges) we have $\fpd(G_N,r)\le 1-\frac1{r+1}$. 
\end{theorem}
In particular, Theorem~\ref{thm:bounded} implies that, under dB updating, the implied scale of fitness of any graph is at most $r+1$. Thus every graph is, at best, a bounded amplifier (see Fig.~\ref{fig:figisf}b). In particular, there exist no linear amplifiers, and thus no quadratic amplifiers or super amplifiers. 
Again, this is in sharp contrast to Bd updating for which super amplifiers exist~\cite{lieberman2005,Giakkoupis16,Galanis17} and, in fact, are abundant~\cite{pavlogiannis2018construction}.

The proof again follows from Lemma~\ref{lem:bound}:
for any $r>1$, the fraction on the right-hand side of Lemma~\ref{lem:bound} is at most the desired $1-1/(r+1)$, with equality when $\d\to\infty$.

\begin{figure}[h] 
  \centering
   \includegraphics[width=0.8\linewidth]{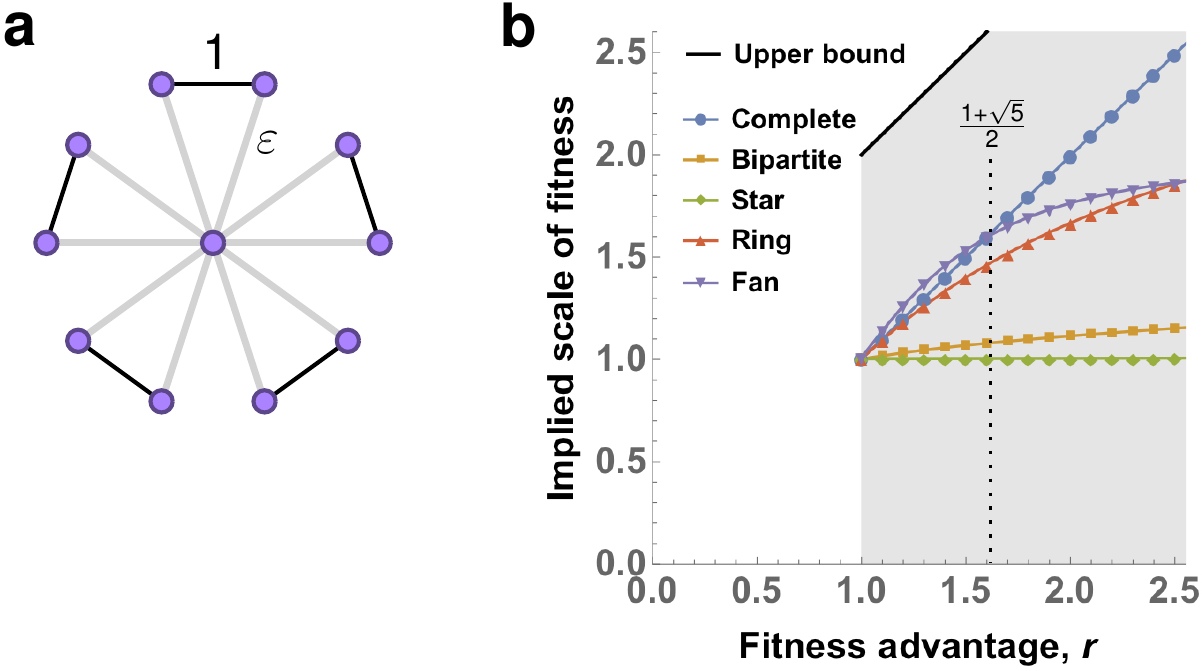}
\caption{ \textbf{Transient amplifiers under death-Birth updating.} A companion work~\cite{BEN} identified certain weighted graphs that are transient dB-amplifiers.
\textbf{a,} The Fan graph $F_{N,\eps}$ with $N$ blades is a weighted graph obtained from a Star graph $S_{2N+1}$ by  pairing up the $2N$ leaves and rescaling the weight of each edge coming from the center to $\eps<1$.
\textbf{b,} The implied scale of fitness of a large Fan (here $N=101$ and $\eps=10^{-5}$).
If $r$ is small enough then the Fan amplifies selection under dB updating. The level of amplification is well within the scope allowed by Theorem~\ref{thm:bounded} (shaded region). For comparison, we again show the implied scale of fitness for the Complete, Bipartite, Star, and Ring graphs ($N=400$). 
}
\label{fig:fans}
\end{figure}

We remark that even though universal amplification is impossible by Theorem~\ref{thm:transient}, some population structures might achieve certain level of amplification for certain range of $r$-values. In fact, a companion work~\cite{BEN} presents weighted population structures called \textit{Fans} that, in the appropriate limit, amplify selection in a range $1<r<(1+\sqrt 5)/2$. 
The extent to which these structures amplify is well within the bounds provided by Theorem~\ref{thm:bounded} (see Fig.~\ref{fig:fans}). It is not known whether there exist unweighted graphs that provide transient amplification.


\subsection*{Extensions to $\delta$-dB amplifiers}
\begin{figure}[ht] 
  \centering
   \includegraphics[width=1\linewidth]{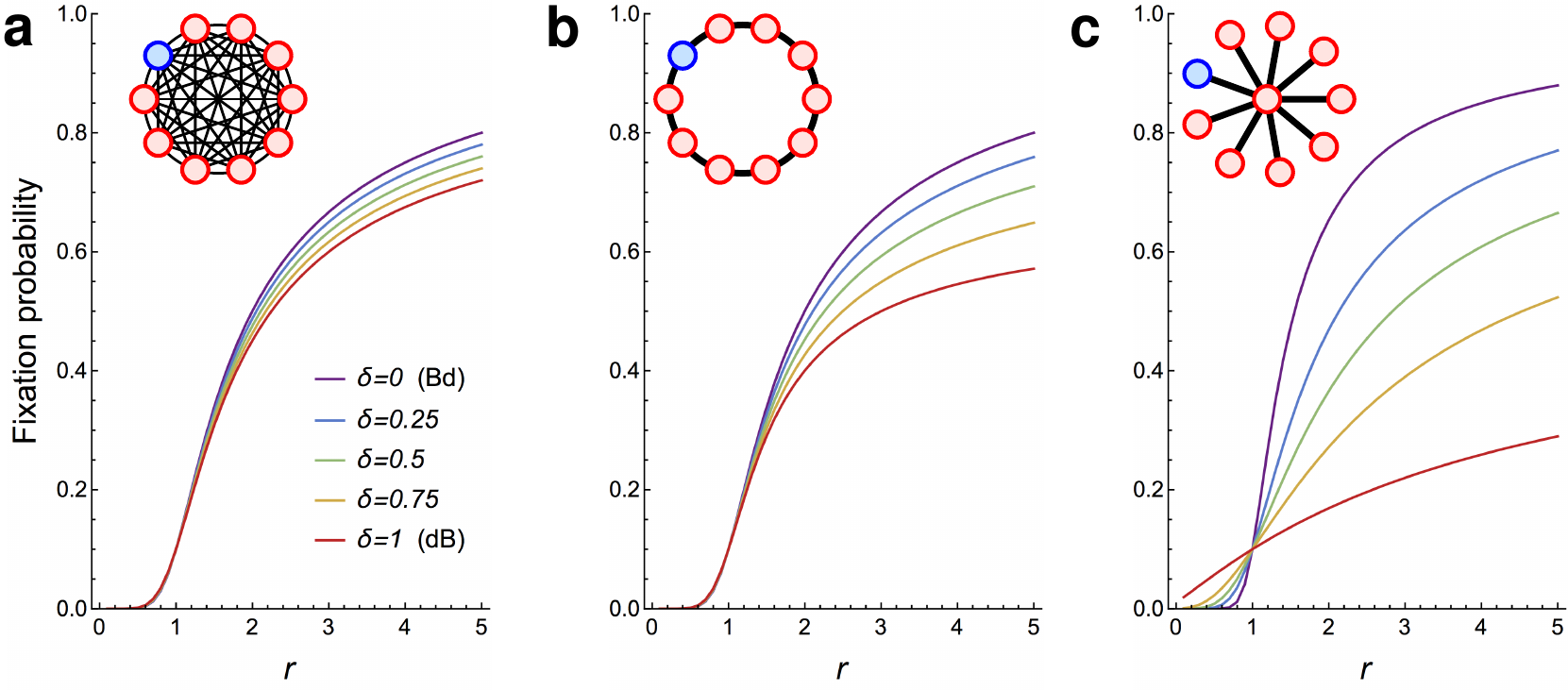}
\caption{ \textbf{Fixation probability under $\delta$-dB updating.}
Three different graphs on $N=10$ vertices: \textbf{a} Complete graph, \textbf{b} Ring graph, \textbf{c} Star graph.
For each $\delta\in\{0,0.25,0.5,0.75,1\}$ we show the fixation probability under $\delta$-dB updating as a function of $r$. On the latter two graphs, the dependence of the fixation probability on $\delta$ is more pronounced and not roughly linear as is the case for the Complete graph. The Star graph is an amplifier under Bd updating and also a $\delta$-dB amplifier for small $\delta$ (e.g. for $\delta=0.2$ and $r=2$ we have $\fpdel(S_{10},r)>0.494>0.491>\fpdelbar(K_{10},r)$) but ceases to be an amplifier for large $\delta$ (e.g. for $\delta=0.5$ and $r=2$ we have $\fpdel(S_{10},r)<0.37<0.47<\fpdelbar(K_{10},r)$).
}
\label{fig:fig3}
\end{figure}

Given the negative answers to questions Q1 and Q2 above, we proceed with studying the $\delta$-dB Moran process, in which the death-Birth updates are interleaved with the Birth-death updates.
The insight of Corollary~\ref{cor:dB_Bd} is that mutants have a higher fixation probability under Bd updating, compared to dB updating (given a large enough fitness advantage $r$).
Qualitatively, we expect that given a fixed population structure under $\delta$-dB updating, the fixation probability increases as $\delta$ decreases.
Fig.~\ref{fig:fig3} confirms this intuition numerically, for Complete graphs, Ring graphs and Star graphs.

The two extremes of the $\delta$-dB Moran process are the pure Bd ($\delta=0$) and pure dB ($\delta=1$) Moran processes.
It is known that under Bd updating, both universal amplifiers and super amplifiers exist.
On the other hand, we have shown here that under pure dB updating, any amplification is inevitably transient and bounded.
The next two natural questions are to investigate whether universal or strong amplifiers exist for small values of $\delta\in(0,1)$, for
which the process is heavily biased towards Bd updating.
Perhaps surprisingly, we answer both questions in negative.
Concerning universality, we show the following theorem.

\begin{theorem}[All $\delta$-dB amplifiers are transient]\label{thm:mixedtransient}
Fix a non-complete graph $G_N$ on $N$ vertices (possibly with directed and/or weighted edges) and $\delta\in(0,1]$. Then there exists $r^\star>1$ such that for all $r>r^\star$ we have $\fpdel(G_N,r) < \fpdelbar(K_N,r)$, where $K_N$ is the complete graph on $N$ vertices.
\end{theorem}

Theorem~\ref{thm:mixedtransient} is a $\delta$-dB analogue of Theorem~\ref{thm:transient}. It implies that, compared to the baseline given by a weighted average $\fpdelbar(K_N,r)$ between $\fpd(K_N,r)$ and $\fp(K_N,r)$, 
every unweighted graph is at best a transient amplifier,
and a weighted graph can only be a universal amplifier if it is a weighted version of the complete graph $K_N$.
Hence for any positive $\delta>0$, no matter how small, universal amplification is impossible among unweighted graphs.

Next, we turn our attention to the limit of large $N$, and ask whether strong amplification is possible for the $\delta$-dB Moran process.
We show the following theorem.

\begin{theorem}[All $\delta$-dB amplifiers are at most linear]\label{thm:mixed}
Fix $r>1$ and $\delta\in(0,1]$. Then for any graph $G_N$ (possibly with directed and/or weighted edges) we have $\fpdel(G_N,r)\le 1- \frac 1{(r/\delta) +1}$.
\end{theorem}

Theorem~\ref{thm:mixed} implies that for fixed $\delta>0$, no matter how small, no better than linear amplifiers exist.
In particular, there are no quadratic amplifiers and no super amplifiers. 
For $\delta\to1$ (pure dB updating), the bound coincides with the one given in Theorem~\ref{thm:bounded}. For $\delta\to0$ (pure Bd updating), the bound becomes vacuous (it simplifies to $\fp(G,r)\le 1$) which is in alignment with the existence of quadratic and super amplifiers under (pure) Bd updating.
The proofs of Theorems~\ref{thm:mixedtransient} and~\ref{thm:mixed} rely on a $\delta$-analogue of Lemma~\ref{lem:bound}.

\begin{figure}[ht] 
  \centering
   \includegraphics[width=1\linewidth]{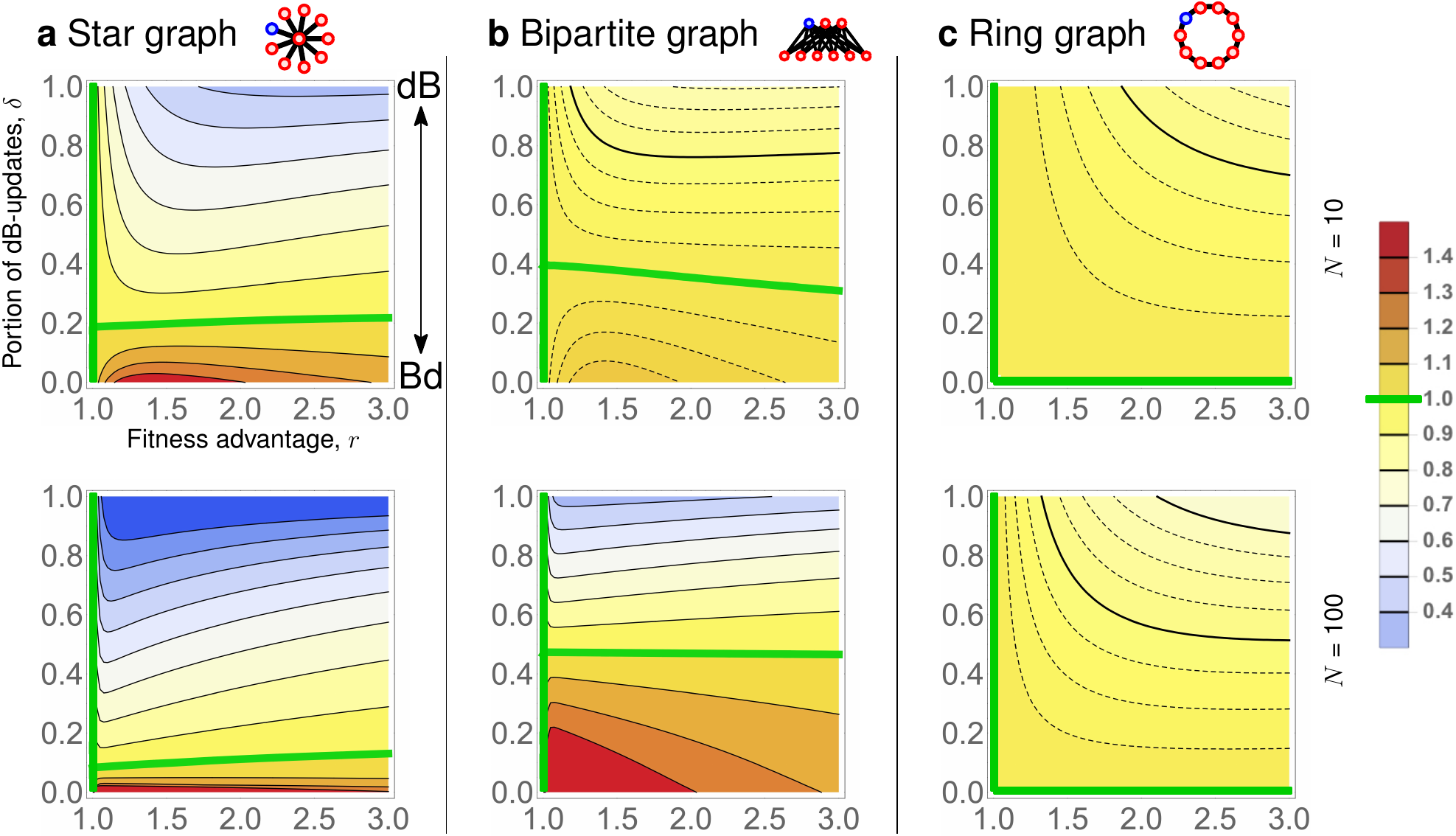}
\caption{ \textbf{Strength of amplification in terms of $r$ and $\delta$.}
\textbf{a}, Star graphs, \textbf{b}, complete Bipartite graphs with smaller part of size $\sqrt{N}$, and \textbf{c}, Ring graphs, of size either $N=10$ (top row) or $N=100$ (bottom row).
For each of the six graphs, we plot the ratio $\fpdel(G_N,r)/\fpdelbar(K_N,r)$ as a function of the fitness advantage $r$ ($x$-axis) and the portion of dB-updates $\delta$ ($y$-axis).
Red (blue) color signifies that the population structure amplifies (suppresses) selection for the given regime $(r,\delta)$. Green curves denote regimes where the ratio equals $1$.
When $r=1$, the fixation probability equals $1/N$ regardless of $\delta$ and the population structure.
By Theorem~\ref{thm:mixedtransient}, all $\delta$-amplifiers are transient, hence the ``horizontal'' green curves eventually hit the $x$-axis for $r$ large enough.
Plotted values were obtained by numerically solving large systems of equations for every $r\in\{1,1.025,\dots,3\}$ and $\delta\in\{0,0.025,\dots,1\}$.
}
\label{fig:fig-delta}
\end{figure}

Even though universal amplification and super amplification are impossible for any $\delta>0$ due to Theorems~\ref{thm:mixedtransient} and~\ref{thm:mixed}, some population structures do achieve reasonable levels of amplification for various combinations of $r$ and $\delta$. Specifically, we consider Star graphs, Bipartite graphs, and Ring graphs of fixed size $N=10$ and $N=100$ and show how strongly they amplify, depending on the fitness advantage $r$ of the initial mutant and on the portion $\delta$ of dB updates (see Fig.~\ref{fig:fig-delta}).
We make several observations.
First, when $\delta$ is small enough, both Star graphs and Bipartite graphs do amplify selection, for a certain range of $r>1$. Interestingly, large Bipartite graphs are less sensitive to variations in $\delta$ than Star graphs, and for small $r>1$ they maintain amplification even for $\delta$ almost as big as $0.5$. On the other hand, if $\delta$ is small enough, Star graphs tend to achieve stronger amplification than Bipartite graphs.
Second, for any of the six population structures and any fixed $r$, increasing $\delta$ diminishes any benefit that the population structure provides to advantageous mutants.
Specifically, there appears to be no regime $(r,\delta)$ where a ring graph would amplify selection.
This is in perfect alignment with Corollary~\ref{cor:dB_Bd}.

\section{Discussion}
In this work, we have investigated the existence of amplifiers for the death-Birth (dB) Moran process.
We have shown that such amplifiers, if they exist, must be both transient and bounded.
Transience means that any population structure can amplify selection only in a limited range $r\in(1,r^\star)$ of relative fitness values $r$ of the mutant.
Boundedness means that even when a population structure does amplify selection for a fixed $r>1$, it can do so only to a limited extent.
In particular, quadratic amplification which is achieved by the Star graphs under Birth-death (Bd) updating is impossible to achieve under dB updating. As a consequence, there are no super amplifiers under dB updating.
These results are in sharp contrast to the Bd Moran process, for which amplifiers and super amplifiers have been constructed repeatedly~\cite{lieberman2005,Monk2014,Galanis17,Goldberg19}, and, in fact, can be abundant~\cite{HT15}.
Our findings suggest that the existence of amplifiers is sensitive to specific mechanisms of the evolutionary process,
and hence their biological realization depends on which process captures actual population dynamics more faithfully.

Note that the situation is more favorable in the broader family of weighted population structures. Under Bd updating, super amplifiers are abundant~\cite{pavlogiannis2018construction}, and under dB updating, transient amplifiers have recently been constructed in a companion work~\cite{BEN}. It remains to be seen whether transient amplification can be achieved by unweighted structures.

To reconcile the apparent discrepancy in the results of the two processes,
we have also investigated the mixed $\delta$-dB Moran process, which combines dB and Bd updating.
On one hand, we have extended our boundedness and transience results to $\delta$-dB updating.
 Specifically, our results imply that for any fixed $\delta>0$, any amplification is necessarily transient and that there are no quadratic amplifiers or super amplifiers under $\delta$-dB updating.
 In this sense, the case of the (pure) Bd updating is singular.
On the other hand, when $\delta$ is small, some population structures that amplify for the pure Bd updating ($\delta=0$) maintain reasonable level of amplification under $\delta$-dB updating, for a wide range of fitness advantages $r$. Specifically, we find that suitable Bipartite graphs are less sensitive to variations in $\delta$ than the Star graphs, and maintain amplification for $\delta$ as big as $0.5$, when $r$ is close to 1.

There is an interesting connection to the situation of evolutionary games on graphs. There, the desirable population structures are those that promote cooperation. It is known that under any $\delta$-dB updating for $\delta>0$, population structures can promote cooperation~\cite{Zukewich13}, whereas for pure Bd updating, no regular structure that promotes cooperation exists~\cite{ohtsuki2006simple}.
Therefore, in the setting of games, the desirable structures exist for all $\delta>0$, whereas in our setting of constant selection, the desirable structures (strong and/or universal amplifiers) exist only in the regime $\delta=0$.
In both settings, the case of pure Birth-death updating appears to be a singular one.

%
%

%
%
\section{Methods}
Here we formally describe the model of Moran process on graphs together with the relevant notions of amplifiers and implied scale of fitness.

\sne{Population structure} In evolutionary graph theory, a population structure is represented by a graph that has $N$ sites (nodes), some of which are connected by 
 edges. Each site is occupied by a single individual. The edge from node $u$ to node $v$ represents that the individual at node $u$ can replace the individual at node $v$. 

\sne{Directions and weights} The edges could be undirected (two-way) or directed (one-way) and they could be weighted. Formally, for a pair of nodes $u$, $v$, the weight of an edge $(u,v)$ is denoted by $w_{u,v}$. If the nodes $u$, $v$ are not connected then $w_{u,v}=0$. In the special case of unweighted graphs, each edge is considered to have weight 1. In the special case of undirected graphs, each edge is two-way. In the most general case of directed graphs with weighted edges, two nodes $u$, $v$ could be interacting in both directions with different weights $w_{u,v}\ne w_{v,u}$. We don't allow self-loops, that is, $w_{u,u}=0$ for each node $u$.

\sne{Mutant initialization} Initially, each site is occupied by a single resident with fitness $1$. Then a single mutant with fitness $r$ appears at a certain node.
This initial mutant node can be selected uniformly at random 
(\textit{uniform initialization})
or with probability proportional to the turnover rate of each node (\textit{temperature initialization}).
Unless specified otherwise, we assume that the initialization is uniform and that the mutation is advantageous ($r>1$).

\sne{Moran dB and Bd updating} Once a mutant has appeared, some version of Moran process takes place. Moran process is a discrete-time stochastic process. At each step, one individual is replaced by a copy of another (neighbouring) individual, hence the population size remains constant.
Denote by $f(v)$ the fitness of the individual at node $v$.
The two prototypical updatings are:
\begin{itemize}
\item Moran death-Birth (dB) updating. An individual $v$ is selected uniformly at random for death. The individuals at the neighbouring sites then compete for the vacant spot. Specifically, once $v$ is fixed, an individual $u$ is selected for placing a copy of itself on $v$ with probability proportional to $f(u)\cdot w_{u,v}$. Note that fitness of an individual doesn't play a role in the death step (thus ``d'' is lower case) but it does play a role in the birth step (thus ``B'' is upper case). 
\item Moran Birth-death (Bd) updating.  An individual $u$ is selected for reproduction with probability proportional to its fitness $f(u)$. Then it replaces a random neighbor. Specifically, once $u$ is fixed, an individual $v$ is replaced by a copy of $u$ with probability proportional to $w_{u,v}$.
\end{itemize}

\sne{Mixed $\delta$-dB updating} The two regimes dB and Bd can be understood as two extreme points of a spectrum. We also consider mixed updating where some steps of the process follow the dB updating while the other ones follow Bd updating. Generally, given a $\delta\in [0,1]$, a $\delta$-dB updating is an update rule in which each step is a dB event with probability $\delta$ and a Bd event with probability $1-\delta$, independently of all the other steps. With this notation, a 1-dB updating is the same as (pure) dB updating and 0-dB updating is the same as (pure) Bd updating.

\sne{Fixation probability} Given a graph $G$, $r>1$ and $\delta\in[0,1]$, we denote by $\fpdel(G,r)$ the fixation probability of a $\delta$-dB updating, when the first mutant is initialized uniformly at random. The complement, that is the probability that the evolutionary trajectory goes extinct, is denoted by $\epdel(G,r)=1-\fpdel(G,r)$.
Specifically, for $\delta=1$ we denote the fixation (resp. extinction) probability under pure dB updating by $\fpd(G,r)$ (resp. $\epd(G,r)$) and similarly for the pure Bd updating which corresponds to $\delta=0$.

\sne{Fixation probability on well-mixed populations}
When studying the effect of population structure on the fixation probability, our baseline is the fixation probability on a well-mixed population of the same size. A well-mixed population is modelled by a complete (unweighted) graph $K_N$, without self-loops. Under pure dB and Bd updating there are exact formulas for fixation probability \cite{kaveh15duality,lieberman2005}:
$$ \fpd(K_N,r)=\frac{N-1}N\cdot \frac{1-\frac1r}{1-\frac1{r^{N-1}}}
  \quad\text{and}\quad
  \fp(K_N,r)=\frac{1-\frac1r}{1-\frac1{r^N}}
  .
$$
For $\delta$-dB updating, no analogous formula is known but numerical computations for various values of $N$ and $r$ show that $\fpdel(K_N,r)$ is essentially indistinguishable from the linear interpolation 
$$\fpdelbar(K_N,r)= \delta\cdot \fpd(K_N,r) + (1-\delta)\cdot \fp(K_N,r)
$$
between $\fpd(K_N,r)$ and $\fp(K_N,r)$ (see Fig. 2 from the main text). 
Therefore, in $\delta$-dB updating we use $\fpdelbar(K_N,r)$ as the baseline comparison.

\sne{Amplifiers of selection} Given $r>1$, some population structures enhance the fixation probability of mutants, compared to the well-mixed population, whereas others decrease it. We refer to the former ones as \textit{amplifiers of selection} and to the latter ones as \textit{suppressors of selection}.
Formally, given a graph $G_N$ with $N$ nodes and some $r>1$, we say that $G_N$ is an \textit{$r$-amplifier under dB updating} if $\fpd(G_N,r)>\fpd(K_N,r)$, where $K_N$ is a complete graph that represents a well-mixed population.
If $G$ is an $r$-amplifier under dB updating for all $r>1$, we call it \textit{universal}. In contrast, graphs that amplify only for some range of values $r\in (1,r^\star)$ are called \textit{transient}.
Similarly, we say that $G_N$ is an \textit{$r$-amplifier under Bd updating} if $\fp(G_N,r)>\fp(K_N,r)$ (note that the baseline is the complete graph $K_N$ under Bd updating) and, for a fixed $\delta\in[0,1]$, we say that $G_N$ is an \textit{$r$-amplifier under $\delta$-dB updating} if $\fpdel(G_N,r)>\fpdelbar(K_N,r)$.

\sne{Classification of amplifiers by strength: Implied scale of fitness} 
Amplifiers can be further classified by strength~\cite{ACN15}. We single out bounded amplifiers, linear amplifiers, quadratic amplifiers and super amplifiers. The intuition behind the classification is that, in the limit of large population size, fixation probability can often be written as $1-1/\isf(r)$ for a suitable function $\isf(r)$ of $r$. For instance, for large well-mixed population we have $\isf(r)=r$ (under any of dB, Bd, $\delta$-dB updating) and for large Star graphs under Bd updating we have $\isf(r)=r^2$. The extent to which a large population structure $G$ distorts this fixation probability can thus be classified by looking at the function $\isf(r)$.

Formally, given a family of graphs $\{G_N\}_{N=1}^{\infty}$ of increasing population size, the \textit{implied scale of fitness} of the family is a function $\isf(r)\colon (1,\infty)\to \mathbb{R}$ such that 
$$\lim\inf_{N\to\infty} \fpd(G_N,r) =1-1/\isf(r).$$

We say that the family is
\begin{enumerate}
\item an (at most) \textit{bounded amplifier} if $\isf(r)\le r+c_0$ for some constant $c_0$.
\item an (at least) \textit{linear amplifier} if $\isf(r)\ge c_1 r +c_0 $ for some constants $c_1>1$, $c_0$.
\item an (at least) \textit{quadratic amplifier} if $\isf(r)\ge c_2 r^2 + c_1r+c_0$ for some constants $c_2>0$, $c_1$, $c_0$.
\item  a \textit{super amplifier} if  $\isf(r)=\infty$ for all $r>1$.
\end{enumerate}
These definitions naturally carry over to Bd updating and $\delta$-dB updating.

\sne{Remark on the regimes considered} We intentionally restrict our attention to the following regimes:
\begin{enumerate}
\item $r>1$. If $r=1$ then $\fpdel(G_N,r)=1/N$, regardless of the population structure. If $r<1$ then $\fpdel(G_N,r)< 1/N\to_{N\to\infty} 0$ for any $G_N$. Thus we focus on $r>1$.
\item Uniform initialization. For dB updating, the notions of uniform and temperature initialization coincide, since every node is, on average, selected for death and replaced equally often. Thus we focus on uniform initialization only.
\item No self-loops. For dB updating, self-loops are not biologically realistic: An individual who has just died can not replace itself. Thus we consider graphs with possibly directed and/or weighted edges but without self-loops.
\end{enumerate}

%
%
\section{Proofs}\label{sec:proofs}
\setcounter{theorem}{0}

Our proofs rely on Jensen's inequality.
For reference purposes, we state it here.
Essentially, given a convex (or concave) function $f$ and several real numbers $x_1,\dots,x_k$, Jensen's inequality bounds the (weighted) average of values $f(x_1),\dots,f(x_k)$ by the value that $f$ takes at the (weighted) average of $x_1,\dots,x_k$.

\begin{claim}[Jensen's inequality]\label{jensen} Let $a_1,\dots,a_n$ be non-negative real numbers that sum up to 1 and let $f$ be a real continuous function. Then
\begin{itemize}
\item If $f$ is convex then
$$ \sum_{i=1}^k a_i\cdot f(x_i)  \ge  f\left(\sum_{i=1}^k a_i\cdot x_i \right)
$$
\item If $f$ is concave then
$$ \sum_{i=1}^k a_i\cdot f(x_i)  \le  f\left(\sum_{i=1}^k a_i\cdot x_i \right)
$$
\end{itemize}
\end{claim}

\subsection{Theorems on dB updating}
The key to proving our theorems on dB updating is the following lemma that gives an upper bound on the fixation probability $\fpd(G,r)$ on an arbitrary graph (possibly with directed and/or weighted edges), in terms of the average in-degree $\d$ and the relative fitness $r>1$ of the mutant.
Recall that given a graph $G$ and its node $v$, the \textit{in-degree} of $v$ is the number of nodes $u$ for which there is an edge $(u,v)$.
If $G$ is undirected then the in-degree of a node is the same as the degree (the number of neighbors).
For any graph $G$, the average in-degree is the same as the average out-degree (and as the average degree if $G$ is undirected).

\lemmadb

\begin{proof}
Denote by $u$ the initial node occupied by the mutant and recall that $\epd(u)$ is the extinction probability under dB updating if the initial mutant appears at $u$. Then $\epd(G,r)=\frac1N\sum_u \epd(u)$.

Denote by $\Ev^-(u)$ (resp. $\Ev^+(u)$) the event that in the next step of the dB updating the number of mutants decreases (resp. increases) and by $p^-(u)$ (resp. $p^+(u)$) the corresponding probability.
Note that if neither of $\Ev^-(u)$, $\Ev^+(u)$ happens, the set of nodes occupied by the mutants stays the same, and if $\Ev^-(u)$ happens before $\Ev^+(u)$, the mutants go extinct. Therefore the extinction probability $\epd(u)$ starting from a configuration with a single mutant at node $u$ satisfies
$$ \epd(u)\ge \frac{p^-(u)}{p^-(u)+p^+(u)}=\frac{1}{1+ \frac{p^+(u)}{p^-(u)} }.
$$

We now compute $p^-(u)$ and $p^+(u)$. The number of mutants decreases if and only if we select the single mutant for death, i.e. $p^-(u)=1/N$, for any node $u$. The number of mutants increases if and only if for death we select some node that neighbors $u$ and then we select $u$ for producing an offspring. Hence 
$$p^+(u)=\sum_v p^+_{u,v},
$$
where
$$p^+_{u,v} = \frac1N\cdot \frac{r\cdot w_{u,v}}{(r-1)\cdot w_{u,v} + \sum_{u'} w_{u',v}}
$$
is the probability that $v$ was selected for death and then $u$ (the only mutant on $G$) was selected to place a copy of itself on $v$.

Now we bound $\ep(G,r)$ in terms of $p^-(u)$ and $p^+(u)$. In the last step we use Jensen's inequality for a function $f(x)=1/(1+x)$ which is convex on $x\in (0,\infty)$:
$$\epd(G,r)=\frac 1N\sum_u \epd(u) \ge \frac1N \sum_u \frac{1}{1+ \frac{p^+(u)}{p^-(u)} } \ge \frac1{1+\frac1N\sum_u  \frac{p^+(u)}{p^-(u)} }.
$$

Since $p^-(u)=1/N$ for all $u$, the right-hand side simplifies and we get
$$\epd(G,r)\ge \frac1{1+\sum_u p^+(u)}.
$$
In the rest, we find a tight upper bound on $\sum_u p^+(u)$.
We first rewrite each $p^+(u)$ using $p^+_{u,v}$ and interchange the sums to get
$$ \sum_u p^+(u) = \sum_u \sum_v p^+_{u,v} = \sum_v \sum_u p^+_{u,v}.
$$
We focus on the inner sum. Fix a node $v$ and denote by $s(v) = \sum_{u'} w_{u',v}$ the total weight of all edges incoming to $v$. Using the formula for $p^+_{u,v}$ we obtain
$$\sum_u p^+_{u,v} = \frac1N \sum_u \frac{r\cdot w_{u,v}}{(r-1)\cdot w_{u,v} + s(v)}.
$$ 
We make three observations.
First, the summation has at most $\din(v)$ terms, where $\din(v)$ is the number of incoming edges to $v$.
Second, we have $\sum_u w_{u,v}=s(v)$.
Third, for fixed $r>0$ and any $s>0$, the function $g(x)=\frac{r\cdot x}{(r-1)x+s}$ is concave on $x\in (0,s)$. Therefore, by another application of Jensen's inequality we can write
$$\sum_u p^+_{u,v} \le \frac1N\cdot \din(v)\cdot \frac{r\cdot \frac{s(v)}{\din(v)}}{(r-1)\cdot\frac{s(v)}{\din(v)} + s(v)} = \frac1N \cdot \frac{r\cdot \din(v)}{r-1+\din(v)},
$$
Finally, summing up over $v$ we obtain
$$\sum_u p^+(u) = \sum_v \sum_u p^+_{u,v} \le \frac1N\sum_v \frac{r\cdot \din(v)}{r-1+\din(v)} \le \frac{r\cdot \d}{r-1+\d},
$$
where in the last step we yet again used Jensen's inequality, this time for the function $h(x)=\frac{r\cdot x}{r-1+x}$ that is concave on $x\in(0,\infty)$, and the fact that the average in-degree of a graph is the same as its average out-degree.

We conclude by observing that this upper bound on $\sum_u p^+(u)$ yields
$$\epd(G,r)\ge  \frac1{1+\sum_u p^+(u)} \ge  \frac1{1+ \frac{r\cdot \d}{r-1+\d}} = \frac{\d+r-1}{\d r+\d+r-1},
$$
hence
$$ \fpd(G,r)\le 1-\epd(G,r)\le \frac{\d\cdot r}{\d\cdot r +\d+r-1}
$$
as desired
\end{proof}

With the lemma at hand, we can prove the first two Theorems.

\thmTransient
\begin{proof} Recall that
$$\fpd(K_N)=(1-1/N)\frac{1-1/r}{1-1/r^{N-1}}\ge (1-1/N)(1-1/r),
$$
hence
$$\epd(K_N)\le \frac{N+r-1}{Nr}.
$$
Using Lemma~\ref{lemma}, it suffices to show that for all sufficiently large $r$ we have 
$$
\frac{\d+r-1}{\d r+\d+r-1} > \frac{N+r-1}{Nr}
$$
which, after clearing the denominators, is equivalent to 
$$r^2\left(N-1-\d \right) -2r(N-1) - (\d-1)(N-1)> 0.
$$
 Since $G$ is not complete, $\d<N-1$ (a strict inequality), hence the coefficient by $r^2$ is positive and the inequality holds for all sufficiently large $r$.
 
 In particular, it is straightforward to check that $r=2N^2$ is large enough: If $G$ misses at least one edge then $\d \le N-1-\frac1N$ hence for $r\ge 2N^2$ the right-hand side is at least
 $$(2N^2)^2\cdot \frac1N - 4N^2(N-1) - N^2 = 3N^2>0.
 $$
\end{proof}

\thmBounded
\begin{proof} Using Lemma~\ref{lemma}, it suffices to check that 
$$  \frac{\d+r-1}{\d r+\d+r-1} \ge \frac1{r+1}
$$
which, after clearing the denominators, is equivalent to $r(r-1)\ge 0$. The equality holds for $r=1$.
\end{proof}

\subsection{Theorems on $\delta$-dB updating}
In order to prove Theorems~\ref{thm:mixed} and~\ref{thm:mixedtransient} we first use a similar technique as before to establish an analogue of Lemma~\ref{lemma} that applies to $\delta$-dB updating.

\lemmadelta

\begin{proof}
 Denote the initial mutant node by $u$ and, as in Lemma~\ref{lemma}, let $p^-(u)$ (resp. $p^+(u)$) be the probability that after a single step of $\delta$-dB updating, the number of mutants in the population decreases (resp. increases).

The values $p^-(u)$ and $p^+(u)$ are weighted averages of the corresponding values under (pure) dB and Bd updating, with weights $\delta$, $1-\delta$. That is,
$$p^-(u)=\ \delta\cdot \frac1N \ \ +\ \  (1-\delta)\cdot \sum_t \frac1{N+r-1}\cdot \frac{w_{t,u}}{\sum_{u'} w_{t,u'}}
$$
and, using the notation $p^+_{u,v}$ from Lemma~\ref{lemma},
$$p^+(u) =\  \delta\cdot \sum_v p^+_{u,v} \ \ +\ \  (1-\delta)\cdot \frac r{N+r-1}.
$$
As in Lemma~\ref{lemma}, we get
$$\epdel(G,r)\ge \frac 1{1+\frac1N \sum_u \frac{p^+(u)}{p^-(u)}}.
$$
For each fixed $u$, we bound $p^-(u)$ from below by ignoring the whole Bd-contribution. We get $p^-(u)\ge \frac \delta N$ which yields
$$\epdel(G,r)\ge \frac 1{1+\frac1\delta \sum_u p^+(u)}
$$
and it remains to bound $\sum_u p^+(u)$ from above.
In $\sum_u p^+(u)$, the total Bd-contribution (summed over $u$) equals $(1-\delta)\frac{Nr}{N+r-1}$
and, as in Lemma~\ref{lemma}, the total dB-contribution is at most $ \delta\cdot \sum_u\sum_v p^+_{u,v}\le\delta\cdot \frac{r\d }{r-1+\d}$.
In total, this yields
$$\epdel(G,r)\ge  \frac{1}{1+  \frac{\d r}{d+r-1} +\frac{1-\delta}{\delta}\cdot \frac{Nr}{N+r-1} }
$$
as desired.
\end{proof}

Using Lemma~\ref{lemma2} we present proofs of Theorems~3 and~4 from the main text. 

\thmMixedTransient
\begin{proof}
Let $d$ be the average in-degree of $G$.
Since $G$ is not complete, we have $d<N-1$ (a strict inequality).

As in the proof of Theorem~\ref{thm:transient}, recall that $\epd(K_N,r)\le\frac{N+r-1}{Nr}$. Moreover, $\fp(K_N,r)=\frac{1-1/r}{1-1/r^N}\ge 1-1/r$, hence $\ep(K_N,r)\le \frac1r$.
This yields
$$1-\fpdelbar(K_N,r)=\epdelbar(K_N,r)= \delta\cdot \epd(K_N,r)+(1-\delta)\ep(K_N,r)\le \frac1r + \delta\cdot\frac{r-1}{Nr}
$$
and by Lemma~\ref{lemma2} it suffices to show that for all sufficiently large $r$ we have
$$ \frac{1}{1+  \frac{\d r}{d+r-1} +\frac{1-\delta}{\delta}\cdot \frac{Nr}{N+r-1} }
\ge \frac1r + \delta\cdot\frac{r-1}{Nr},
$$
Since $N$, $d$ and $\delta$ are all fixed, we can consider both sides as functions of $r$. As $r\to\infty$, the left-hand side tends to $\frac1{1+d+\frac{1-\delta}\delta N}$ while the right-hand side tends to $\frac\delta N$. In order to conclude, it suffices to show strict inequality between the respective limits:
$$\frac1{1+d+\frac{1-\delta}\delta N} > \frac\delta N.
$$
After clearing the denominators, this is equivalent to $\delta(N-1-d)>0$ which indeed holds for any $\delta>0$ and any non-complete graph $K_N$.
\end{proof}

\thmMixed

\begin{proof}
Since $\d\le N-1<N$ and $r>1$, we have
$$ \frac{\d r }{\d+r-1} <  \frac{Nr}{N+r-1},
$$
hence Lemma~\ref{lemma2} gives
$$\epdel(G,r)\ge 
 \frac{1}{1+  \frac{\d r}{d+r-1} +\frac{1-\delta}{\delta}\cdot \frac{Nr}{N+r-1} }
 > \frac 1{1+\frac1\delta\cdot\frac{Nr}{N+r-1}} \ge \frac{\delta}{r+\delta} = \frac1{(r/\delta)+1}, 
$$
where the last inequality is equivalent to $\delta\cdot r(r-1)\ge 0$ after clearing the denominators.
The result follows.
\end{proof}

%
%
\section{Further directions}\label{sec:questions}
\begin{figure}[ht] 
  \centering
   \includegraphics[width=0.6\linewidth]{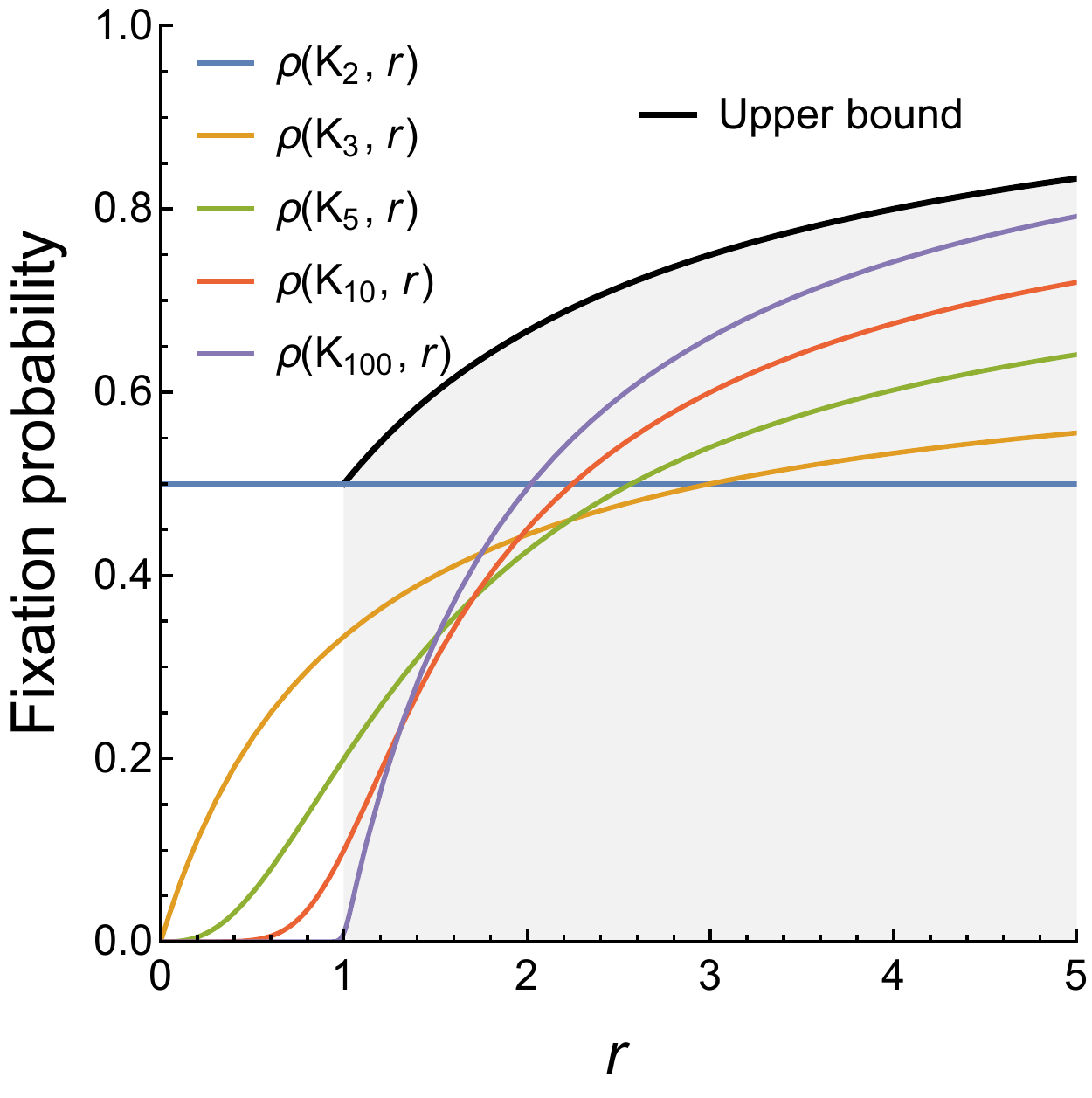}
\caption{ \textbf{Additional figure: Tightness of the upper bound.}
We consider Complete graphs of sizes $N\in\{2,3,5,10,100\}$ under dB updating. The fixation probability is always below the upper bound given by Theorem~\ref{thm:bounded}. For $r=1$ the bound precisely matches the fixation probability on $K_2$. For large $r$, the bound is relatively tight with respect to large Complete graphs.
}
\label{fig:additional}
\end{figure}

Here we list several interesting open questions.
\begin{enumerate}
\item \textbf{Unweighted transient amplifiers under dB updating.} Weighted transient amplifiers for dB updating have been constructed in a companion work~\cite{BEN}. Do there exist transient amplifiers among unweighted graphs?

\item \textbf{Towards universal amplification under dB updating.}
The weighted transient amplifiers constructed in the companion work~\cite{BEN} amplify for $r$ less than a golden ratio $\phi=\frac12(1+\sqrt 5)\doteq 1.618$. Does there exist a graph that amplifies for $r>\phi$? If so, does for every $r^\star$ exist a graph that amplifies for all $r\in(1,r^\star)$? If so, does there exist a universal amplifier for dB updating? Theorem~\ref{thm:transient} implies that if so, it has to be a weighted version of a complete graph. 

\item \textbf{Towards universal amplification under $\delta$-dB updating.} How do the answers change under $\delta$-dB updating instead of (pure) dB updating? Specifically, do large complete Bipartite graphs amplify on arbitrarily large intervals $(1,r^\star)$, provided that $\delta$ is small enough? 

\item \textbf{Well-mixed populations with $\delta$-updating.} Is there a simple formula for fixation probability on a 
complete graph under $\delta$-dB updating for $\delta\in(0,1)$?

\item \textbf{Monotonicity in $\delta$.} Is $\fpd(G_N,r)<\fp(G_N,r)$ for any fixed graph $G$ and any fixed $r>1$? If so, is $\fpdel(G_N,r)$ a decreasing function of $\delta$, for any fixed graph $G$ and any fixed $r>1$?

\item \textbf{Optimal graph for a given $r$.} For fixed $r>1$, what is the highest possible fixation probability $\fpd(G,r)$, attained by any graph $G$?
Theorem~\ref{thm:bounded} states that $\fpd(G,r)\le 1-1/(r+1)$ for any fixed $r>1$ and any graph $G$. The bound is attained for $r=1$ due to $K_2$ and is relatively tight for $r\to\infty$ due to large Complete graphs which give $\fpd(K_N,r)\to_{N\to\infty} 1-1/r$ (see Figure~\ref{fig:additional}). Are those graphs optimal? Or does there exist $r>1$ and a graph $G$ (of any size) such that 
$\fpd(G,r)> \max\{\frac12, 1-\frac1r\}?$
\end{enumerate}

\section*{Acknowledgments}
J.T. and K.C. acknowledge support from ERC Start grant no. (279307: Graph Games), Austrian Science Fund (FWF) grant no. P23499-N23 and S11407-N23 (RiSE).
A.P. acknowledges support from FWF Grant No. J-4220.
M.A.N. acknowledges support from Office of Naval Research grant N00014-16-1-2914 and from the John Templeton Foundation. 
The Program for Evolutionary Dynamics is supported in part by a gift from B. Wu and E. Larson.

\bibliography{refs}

\end{document}